\documentstyle[12pt]{article}

\newcommand{\be}{\begin{equation}}
\newcommand{\ee}{\end{equation}}
\newcommand{\bea}{\begin{eqnarray}}
\newcommand{\eea}{\end{eqnarray}}

\begin{document}
\begin{titlepage}

\flushright{IP-BBSR-2010-03 }

\vspace{1in}

\begin{center}
\Large
{\bf MASSIVE STRINGY STATES AND T-DUALITY  SYMMETRY}

\vspace{1in}

\normalsize

\large{  Jnanadeva  Maharana\\
E-mail: maharana$@$iopb.res.in \\
(\today )}

\normalsize
\vspace{.5in}

 {\em Institute of Physics \\
Bhubaneswar - 751005 \\
India  \\ }

\end{center}

\vspace{1in}

\baselineskip=24pt
\begin{abstract}
We present evidence for the target space duality symmetry associated with
massive excited states of closed bosonic string. The evolution of string is
considered in ${\hat D}$ spacetime dimensions; out of which 
$d$ spacial dimensions are compactified
on torus.
The phase space Hamiltonian
formulation is adopted to unveil the T-duality symmetry from the worldsheet
perspective. The existence of this symmetry is verified for a few massive
levels. 
A systematic procedure is presented to
study T-duality symmetry of vertex operators for all massive levels of
closed bosonic string. It is argued that all vertex operators corresponding
to  excited massive states can be cast in an $O(d,d)$ invariant form where
$d$ is the number of compact dimensions.

\end{abstract}

\vspace{.7in}

\end{titlepage}



The string theory has emerged as the most promising candidate to
 unify the fundamental forces of Nature . It is recognized that several
important developments have taken place in this direction \cite{books}.
Some of the crucial  issues pertaining to quantum 
gravity have been
addressed in the framework of string theory with considerable success. 
The computation of
Bekenstein-Hawking entropy and insight into the nature of Hawking radiation
in black hole physics from a microscopic theory are considered as important
achievements. Moreover, outcome of important results of string theory 
 have influenced research in the frontiers of cosmology.
 There has been considerable progress to build
realistic models in string theory framework in order to provide a
deeper understanding of the fundamental interactions, although the standard model
of particle physics has not emerged as yet. The  
string dynamics, in its first quantized formulation, is primarily guided by 
duality symmetries \cite{duality}.  
The web of dualities unravel the intimate relationships
between the five string theories in various dimensions \cite{ht,w}
although they are
perturbatively distinct in the critical dimension, ${\hat D}=10$. The target
space duality, T-duality, is a special attribute of the theory 
which owes its origin to
the one dimensional nature of the string. In its simplest form, we encounter
T-duality in the worldsheet description of string's evolution. If $\sigma$
denotes its coordinate, in the temporal evolution ($\tau$-evolution), a string 
sweeps a surface and therefore,  its coordinates, 
$X^{\hat\mu}(\sigma,\tau)$ are
 parametrized by them. Thus $\tau \leftrightarrow \sigma$ interchange describes
the same physical evolution process. When particles, which belong to the string
spectrum, scatter they form strings as intermediate states corresponding to
exchange of towers of particles. Interchange of $\tau ~ {\rm and}~ \sigma$
is to be interpreted as direct channel and crossed channel processes from
the quantum field theory perspective.\\ 
The purpose of this note is to explore duality symmetry associated with massive
excited stated of closed string in the Hamiltonian formalism from the 
worldsheet view point. The evolution of the string in the background of its 
massless excitation corresponds to a 2-dimensional $\sigma$-model where the
backgrounds are identified as coupling constants of the theory. These are
constrained if we demand that the theory respects conformal
invariance. We intend to follow a similar approach where the string evolves
in the background of higher massive  levels 
in order to study the
duality symmetry associated with the excited states. \\
Let us recall that all the massive states of closed string belong to 
irreducible representations of rotation group, $SO({\hat D}-1)$, $\hat D$
being the number of spacetime dimensions. Moreover,  at each level of the
spectrum, the states are degenerate  with different spins. 
The importance of massive string states is recognized when one computes
the $\beta$-functions associated with the massless states, in the
$\sigma$-model approach, beyond the leading order. In particular, when one
computes the second order corrections to the $\beta$-function in the graviton 
background, it was observed that \cite{bala1,itoi} it is necessary to introduce
counter terms which correspond to infinite number of massive modes to cancel
the loop diverges that appear in the bosonic theory. 
One could couple the string to excited massive states
analogously, generalizing the procedures of coupling to the massless 
excitations; however, such terms will be suppressed by appropriate factors
$\alpha '$ on
dimensional considerations. Therefore, the resulting effective actions
(obtained for such massive states) will play a subdominant role in the low
energy regime. Let us recall that compactification of string effective action,
derived for a string in critical dimensions(26 and 10 for bosonic and
superstrings respectively), leads to appearance of noncompact duality group
(see \cite{ms} and references therein).
Thus it is worth while to examine whether excited states are endowed with
any duality-like symmetry. We conjecture that there are evidences for dualities
which can be verified, for toroidally compactified closed bosonic string,
in the case of  a few excited levels. Moreover, we present a systematic
procedure to verify the validity of our proposal that  
 such dualities persist for all massive levels. 
It is well known that
the excited states are endowed with several interesting
attributes. The degeneracy of higher states, for a given mass,
 grows exponentially which is the {\it raison de etre} for limiting (Hagedorn) 
temperature. 
Moreover, it has been argued that in the Planckian energy scattering regime,
such stringy states play an important role
\cite{planck1,planck2}. There are hints that these states
might be endowed with higher gauge symmetries \cite{mv,ov1,ov2,ao,mm,laba}. 
The recent interests to study high spin massless field theories have utilized
properties of excited stringy states in certain limits \cite{sagnoti,others}.
We recall some of the useful results in order to formulate our problem.
In this optics, it is worth while to investigate duality symmetry associated 
with excited massive levels of closed string where $d$ of its spatial
coordinates are compactified on $T^d$.\\
The worldsheet action for a closed bosonic string evolving in flat Minkowski
target space is given by
\bea
\label{freeaction}
S={1\over 2}\int d\sigma d\tau \bigg({\gamma^{ab}}
G^{(0)}_{{\hat \mu}{\hat{\nu}}}\partial_a X^{\hat{\nu}}\partial 
X^{\hat{\nu}}+\epsilon^{ab}B^{(0)}_{{\hat \mu}{\hat{\nu}}}
\partial_a X^{\hat{\nu}}\partial X^{\hat{\nu}}    \bigg)
\eea
where $\gamma^{ab}={\rm diag}(1,-1)$ worldsheet metric in the orthonormal 
gauge, ${\hat\mu},{\hat\nu}=0,1,2,... {\hat D}$, 
 $G^{(0)}_{{\hat\mu}{\hat\nu}}={\rm diag} (1,-1,-1...)$ is the 
Minkowski metric of the target space, $B^{(0)}_{{\hat \mu}{\hat{\nu}}}$
is the constant antisymmetric tensor; consequently, the presence of the last term in
(\ref{freeaction}) does not contribute to the equations of motion.
When $G_{{\hat\mu}{\hat\nu}}$ is a constant metric
the perturbative spectrum remains invariant under $G^{(0)}
\leftrightarrow G^{^{-1}(0)}$ and $P_{{\hat\mu}}
\leftrightarrow X'^{{\hat\mu}}$ and the corresponding 
Hamiltonian 
also remains invariant under this duality. If we couple the string to spacetime 
dependent backgrounds 
$G_{{\hat\mu}{\hat\nu}}(X)~{\rm and}~B_{{\hat\mu}{\hat\nu}}(X)$
conformal invariance imposes strong constraints on their admissible 
configurations through differential equations; the so called equations of
motion. These equations assume covariant
forms if we adopt the well known technique of Riemann normal coordinate
expansion method. On the other hand if one assumes these backgrounds to be
weak ($G_{{\hat\mu}{\hat\nu}}(X)\sim G^{(0)}_{{\hat\mu}{\hat\nu}}+h_{{\hat\mu}
{\hat\nu}}(X)$ and $ B_{{\hat\mu}{\hat\nu}}(X) \sim b_{{\hat\mu}{\hat\nu}}(X)$)
and demands conformal  invariance on the vertex operators 
then these massless backgrounds
satisfy transversality condition and equations of motion. One of the most
powerful and efficient techniques is to compute their conformal weights
with respect to the worldsheet stress energy momentum tensor obtained from
(\ref{freeaction}) and require them to be $(1,1)$ primaries. Alternatively,
we can adopt the BRST formalism which also yields the same constraints.\\
Our principal goal is to investigate the duality properties of massive excited
string states in the weak field approximation. We assume that the target space
metric is flat and set $ B_{{\hat\mu}{\hat\nu}}=0$. 
Thus the Hamiltonian, we are going to deal with,  is
the sum of the free Hamiltonian obtained from (\ref{freeaction}) and
the terms coming from vertex operators of massive levels.  
If we require the vertex operators to be $(1,1)$ primaries \cite{ov1,ov2,ao}
then they satisfy equations of motion as well as some transversality 
conditions. We shall not provide details of these calculations which
are available in the literature \cite{ov1,ov2}, although we shall
utilize them when the need arises. The  target space duality has been
examined from several perspectives \cite{before}.
Now we recall
some salient results of T-duality in the frame work of the worldsheet
theory \cite{duff,jm,ms,siegel}. In particular we focus on
 toroidal compactification for massless states in the worldsheet approach
\cite{duff,jm,ms}.
In this
context it is assumed that string coordinates 
$Y^{\alpha}(\sigma ,\tau),\alpha , \beta=1,2,..d$ are compactified 
on torus $T^d$. 
The noncompact coordinates are
$X^{\mu}(\sigma,\tau), \mu ,\nu=0,1,2..D-1$ with $D+d={\hat D}$. The 
corresponding backgrounds after dimensional reduction\cite{ms}, for the
  metric, are 
$G_{\mu\nu}(X), A^{(1)}_{\mu\alpha}(X) {\rm and}~G_{\alpha\beta}(X)$ and 
from the 2-form we get 
$B_{\mu\nu}(X), B_{\mu\alpha} ~{\rm and}~B_{\alpha\beta}(X)$.
It is assumed that all the backgrounds depend only on the 
spacetime string coordinates 
$X^{\mu}$. The gauge fields $A^{(1)}_{\mu\alpha}$ are 
associated with the
isometries and $B_{\mu\alpha}$ are another set of gauge fields coming from
dimensional reductions of the 2-form. It was shown that, after introducing
a set of dual coordinates ${\tilde Y}^{\alpha}$ the combined 
worldsheet equations of 
motion (of $Y ~{\rm and} ~ {\tilde Y}$) can be cast in a duality covariant form.
Note  that if one resorts to the  Hamiltonian formulation for a slightly
simplified version of above compactification \cite{hs}, 
the resulting Hamiltonian is expressed in duality invariant form \cite{jmodd}.
 Our strategy
will be to utilize the results of Hamiltonian formulation and adopt a
simple compactification procedure for the higher levels  and unveil the duality
symmetry for these states. Let us consider toroidal compactification where
we set $G_{\alpha\beta}=\delta _{\alpha\beta} ~{\rm and}~ B_{\alpha\beta}=0$;
in other words the radii of $T^d$ are set to unity. The stress energy momentum
tensors used to compute the conformal weights are
\bea
\label{tensors}
T_{++}={1\over 2}(G^{(0)}_{\mu\nu}\partial X^{\mu}\partial X^{\nu}+
\delta _{\alpha\beta}\partial Y^{\alpha}\partial Y^{\beta} )
\eea 
and
\bea
\label{tensor2}
T_{--}={1\over 2} (G^{(0)}_{\mu\nu}{\bar\partial}X^{\mu}{\bar\partial}X^{\nu} +
{\bar\partial}Y^{\alpha}{\bar\partial}Y^{\beta})
\eea
where $G^{(0)}_{\mu\nu}={\rm diag}(1,-1,-1..)$ is the flat D-dimensional metric,
$\partial X^{\mu}={\dot X}^{\mu}+X'^{\mu}$,
$\partial Y^{\alpha} ={\dot Y}^{\alpha}+Y'^{\alpha}$, 
${\bar\partial} X^{\mu}={\dot X}^{\mu}-X'^{\mu}$ and
${\bar\partial}Y^{\alpha}= {\dot Y}^{\alpha}-Y'^{\alpha}$; 
'overdot' and 'prime' stand for derivatives with  
respect to $\tau$ and $\sigma$ here and everywhere. 
Note that the string coordinates,
$X^{\mu}$ and $Y^{\alpha}$ will be decomposed as right and left movers,
$X^{\mu}=X_R^{\mu}+X_L^{\mu}, ~{\rm and }~ Y^{\alpha}=Y_R^{\alpha}+Y_L^{\alpha}$ 
and therefore, 
$\partial {\bar\partial} X^{\mu}= \partial {\bar\partial} Y^{\alpha}=0.$
Therefore, we do not include such terms in construction of excited level vertex
operators. The vertex operator for first excited massive state \cite{ov1,ov2} for the
uncompactified closed string is 
\bea
\label{phione}
{\bf{\hat{\Phi}}_1}= {\hat V}^{(1)}_1 +{\hat V}^{(2)}_1+ {\hat V}^{(3)}_1
\eea
where
\bea
\label{vetex1}
{\hat V}^{(1)}_1 = A^{(1)}_{{\hat\mu}{\hat\nu} ,{\hat\mu}'{\hat\nu}'}(X)
\partial X^{\hat\mu}\partial X^{\hat\nu}{\bar\partial}X^{{\hat\mu}'}
{\bar\partial}X^{{\hat \nu}'} 
\eea
\bea
\label{vertex2}
{\hat V}^{(2)}_1= A^{(2)}_{{\hat\mu}{\hat\nu},{\hat\mu}'}(X)
\partial X^{\hat\mu}\partial X^{\hat\nu}{\bar\partial}^2X^{{\hat\mu}'},~~~
{\hat V}^{(3)}_1= A^{(3)}_{{\hat\mu},{\hat\mu}'{\hat\nu}'}(X)
\partial ^2 X^{\hat\mu}
{\bar\partial}X^{{\hat\mu}'}{\bar\partial}X^{{\hat \nu}'}
\eea
\bea
\label{vertex3}
{\hat V}^{(4)}_1=A^{(4)}_{{\hat\mu},{\hat\mu}'}(X)\partial ^2 X^{\hat\mu}
{\bar\partial}^2X^{{\hat\mu}'}
\eea
The subscript '1' appearing in ${\hat V}^{(i)}_1, i=1-4$ is indicative of the fact
that they correspond to ones for the first excited massive
level.
Notices that the tensor indices are labeled with unprimed and primed indices.
This convention is adopted to keep track of the operators (or oscillators
in mode expansions of $X^{\hat\mu}$) coming from the right moving sector such
as $\partial X^{\hat\mu}$ and from the left moving sector, 
${\bar\partial} X^{{\hat\mu}'}$, or powers of $\partial,~ {\bar\partial}$
acting on $X^{\hat\mu}$. It facilitates our future
computation and will be useful 
notation when we dwell on duality symmetry in sequel. If we demand 
${\bf{\hat{\Phi}}_1}$ to be a $(1,1)$ primary, with respect to $T_{\pm\pm}$, 
then
 $V^{(i)}_1$ are are constrained (actually  the 
$X^{\hat\mu}$-dependent tensors,$A^{(i)}$ are restricted). 
It is a straight forward calculation to obtain these conditions.  We follow
the methods of \cite{ov1,ov2} and summarize the relevant results below. These
will be utilized when we explore the associated  T-duality properties of
these vertex operators for the compactified scenario. Note that each one of the 
functions, ($V^{(2)}_1-V^{(4)}_1$),  is not $(1,1)$ on its own; however,
$V^{(1)}$ is $(1,1)$ as is easily verified. Second point, 
we mention in passing, 
is that conformal invariance imposes two types of constraints on these vertex 
functions. We designate  $A^{(i)}_1$ or $V^{(i)}_1$ as vertex functions to
distinguish them 
from full vertex operator for a given level, like ${\bf{\hat\Phi _1}}$
for the first excited sate, which is expressed as sum of vertex functions.
: each one satisfies a mass-shell condition (recall that same is true
for tachyon and all massless vertex operators) and gauge (or transversality)
conditions which is also known for all the massless sectors. These are listed
below
\bea
\label{shell1}
({\hat\nabla }^2-2)A^{(1)}_{{\hat\mu}{\hat\nu} ,{\hat\mu}'{\hat\nu}'}(X)=0,~~
({\hat\nabla} ^2-2) A^{(2)}_{{\hat\mu}{\hat\nu},{\hat\mu}'}(X)=0, 
\eea
and
\bea
\label{shell2}
({\hat\nabla} ^2-2) A^{(3)}_{{\hat\mu},{\hat\mu}'{\hat\nu}'}(X)=0,~~
({\hat\nabla} ^2-2)A^{(4)}_{{\hat\mu},{\hat\mu}'}(X)=0
\eea
The ${\hat D}$-dimensional laplacian, ${\hat\nabla}^2$, is 
defined in term of the
flat spacetime metric. The mass levels are in in units of the string scale
which has been set to {\it one} in eqs.(\ref{shell1}) and (\ref{shell2}).
The four vertex functions are also related through following equations
\bea
\label{relation}
A^{(2)}_{{\hat\mu}{\hat\nu} ,{{\hat\mu}'}}=
\partial^{{\hat\nu}'}A^{(1)}_{{\hat\mu}{\hat\nu},{{\hat\mu}'}{{\hat\nu}'}},~~~
A^{(3)}_{{\hat\mu},{{\hat\mu}'}{{\hat\nu}'}}=\partial^{\hat\nu} 
A^{(1)}_{{\hat\mu}{\hat\nu} ,{\hat\mu}'{\hat\nu}'},~~~
A^{(4)}_{{\hat\mu},{\hat\mu}'}=\partial^{{\hat\nu}'}\partial^{{\hat\nu}}
A^{(1)}_{{\hat\mu}{\hat\nu} ,{\hat\mu}'{\hat\nu}'}
\eea
Here $\partial^{{\hat\mu}}$ etc. stand for partial derivatives with respect to
spacetime coordinates. Furthermore, besides eqs. (\ref{shell1}),(\ref{shell2}) 
and eq. (\ref{relation}) there are further constraints  (like gauge conditions)  
which also follow from
the requirements of that the vertex functions be $(1,1)$ primaries 
\cite{ov1,ov2} 
\bea
\label{gauge}
{A^{(1){\hat\mu}}_{\hat\mu}},_{{\hat\mu}'{\hat\nu}'}+2\partial^{\hat\mu}
\partial^{\hat\nu}A^{(1)}_{{\hat\mu}{\hat\nu},{{\hat\mu}'}{{\hat\nu}'}}=0,~~~
{\rm and}~~~{A^{(1)}_{{\hat\mu}{\hat\nu},{{\hat\mu}'} }} ^{{\hat\mu}'}+
2\partial^{{\hat\mu}'}
\partial^{{\hat\nu}'}A^{(1)}_{{\hat\mu}{\hat\nu},{{\hat\mu}'}{{\hat\nu}'}}=0
\eea
The above relations, eq.(\ref{relation}) and eq.(\ref{gauge}), will be useful 
for our investigation of the duality in what follows.\\
Let us very briefly recapitulate how the T-duality group $O(d,d)$ plays
an important role in the worldsheet Hamiltonian description of a closed 
string compactified on $T^d$. We shall proceed in two steps.
First consider a simple compactification scheme \cite{hs}where the 
metric and and the 2-form decompose as follows
\bea
\label{decompose}
G_{\hat\mu\hat\nu}(X)=\pmatrix{g_{\mu\nu}(X) & 0\cr 0 &
G_{\alpha\beta}(X) \cr},~~
B_{\hat\mu\hat\nu}=\pmatrix{b_{\mu\nu}(X) & 0 \cr 0 &
B_{\alpha\beta}(X) \cr}
\eea
We shall consider the  following action. We intend to go  over 
to canonical Hamiltonian 
description;  note that  $\gamma ^{ab}$ is already chosen to be ON gauge metric).
\bea
\label{action}
S={1\over 2}\int d\sigma d\tau \bigg( \gamma ^{ab}{\sqrt {-\gamma}}
G_{\hat\mu\hat\nu}(X)\partial_a X^{{\hat\mu}}\partial X^{{\hat\nu}}+
\epsilon ^{ab}B_{\hat\mu\hat\nu}(X)
\partial _a X^{\hat\mu}\partial _b X^{\hat\nu} \bigg)
\eea
We introduce a pair of vectors ${\cal V}$ and ${\cal W}$ of dimensions
$2D$ and $2d$ respectively as defined below
\bea
\label{vectors}
{\cal V}=\pmatrix{{\tilde P}_{\mu} \cr X'^{\mu}\cr},~~
{\cal W}=\pmatrix{{ P}_{\alpha}\cr Y'^{\alpha}\cr}
\eea
where ${\tilde P}_{\mu} ~{\rm and}~ P_{\alpha}$ are conjugate momenta of string
coordinates $X^{\mu} ~{\rm and}~ Y^{\alpha}$ respectively. The canonical
Hamiltonian is expressed as sum of two terms
\bea
\label{newh}
{\cal H}_c={1\over 2}\bigg({\cal V}^T{\tilde M}{\cal V}+
{\cal W}^T{ M}{\cal W}\bigg)
\eea
whereas ${\tilde M}$ is a  $2D\times 2D$ matrix and $ M$ is another
 $2d\times 2d$ matrix, given by
\bea
\label{newmatrix}
{\tilde M}=\pmatrix{g^{\mu\nu} & -g^{\mu\rho}b_{\rho\nu} \cr
b_{\mu\rho}g^{\rho\nu} &
g_{\mu\nu}-b_{\mu\rho}g^{\rho\lambda}b_{\lambda\nu} \cr}, ~~
{ M}=\pmatrix{G^{\alpha\beta} & -G^{\alpha\gamma}B_{\gamma\beta} \cr
 B_{\alpha\gamma}G^{\gamma\beta} & G_{\alpha\beta}-B_{\alpha\gamma}
G^{\gamma\delta}B_{\delta\beta} \cr}
\eea
Let us focus on the second term of (\ref{newh}), define it to be 
$H_2={1\over 2}{\cal W}^TM{\cal W}$,
 which is of importance to us.  Under the global $O(d,d)$ 
transformations
\bea
\label{odd2}
{ M}\rightarrow \Omega{ M}\Omega ^T,~
{\cal W}\rightarrow \Omega {\cal W},~ \Omega ^T{\bf\eta}\Omega ={\bf\eta},~
\Omega \in O(d,d),~ {\bf\eta}=\pmatrix{0 & {\bf 1} \cr {\bf 1} & 0 \cr}
\eea
$\bf \eta$ is the $O(d,d)$ metric and $\bf 1$ is $d\times d$ unit matrix
 and $\cal W$ is the $O(d,d)$ vector and $M\in O(d,d)$.
 Since, $\tilde M$ and
$\cal V$ are {\it inert} under this duality transformation; as a consequence, 
${\cal H}_c$ is indeed T-duality invariant. The moduli, ${\tilde M}$ and  $M$
are classical backgrounds. However, we shall work in the weak field 
approximation: $G_{\alpha\beta}=\delta_{\alpha\beta}+h_{\alpha\beta}$. 
Let us focus on the $O(d,d)$ invariance of graviton 
vertex operator 
along compact directions
\bea
\label{vhvertex}
V_h=h_{\alpha{\beta}'}\partial Y^{\alpha}{\bar \partial}Y^{\beta'}
\eea
$h_{\alpha{\beta}'}$ is a symmetric tensor. Noting that  
$P_{\alpha}=\delta_{\alpha\beta}{\dot Y}^{\alpha}$, we rewrite 
(\ref{vhvertex}) as 
\bea
\label{gravvert}
V_h=h^{\alpha{\beta}'}P_{\alpha}P_{\beta'}-
h_{\alpha{\beta}'}Y'^{\alpha}Y'^{{\beta}'}
\eea
This is expressed in $O(d,d)$ invariant form
\bea\label{gravodd}
V_h={\cal W}^T{\bf H}{\cal W},~~ {\bf H}=\pmatrix{h & 0\cr 0 &-h \cr}
\eea
${\bf H} \in O(d,d)$ and transforms according to (\ref{odd2}); the 
appropriate assignment of indices can be read off from (\ref{gravvert}). We can
repeat the same procedure for the combined vertex of graviton and antisymmetric
tensor, $b_{\alpha\beta'}$. \\
Let us examine T-duality properties of the first excited massive level where
we adopt a simple compactification scheme. 
We focus the attention on $V^{(1)}_1$
as an example. Note that if we follow the toroidal comactification scheme
adopted in \cite{ms} in the context of worldsheet duality for the case at
hand the vertex function $A^{(1)}_{{\hat\mu}{\hat\nu} ,{\hat\mu}'{\hat\nu}'}(X)
\partial X^{\hat\mu}\partial X^{\hat\nu}{\bar\partial}X^{{\hat\mu}'}
{\bar\partial}X^{{\hat \nu}'}$ will decompose into following forms: (i)
A tensor $A^{(1)}_{\mu\nu ,\mu '\nu '}$, one which has all Lorentz indices
(ii) another which has three Lorentz indices and one index corresponding to
compact directions, (iii) a tensor with two Lorentz indices and two indices
in compact directions, (iv) another,  which has a single Lorentz index and
three indices in in internal directions and (v) a  tensor with all indices
corresponding to compact directions i.e. 
$A^{(1)}_{\alpha\beta ,\alpha '\beta '}$. It is obvious these tensors will be
suitably contracted with $\partial X^{\mu}, {\bar\partial}X^{\mu},
\partial Y^{\alpha}, {\bar\partial}Y^{\alpha}$ with all allowed combinations.
We adopt, to start with,  a compactification scheme where only 
$A^{(1)}_{\alpha\beta ,\alpha '\beta '}$ is present and the tensors with mixed
indices are absent. We shall return to more general case later. 
We may allow the presence of 
$A^{(1)}_{\mu\nu ,\mu '\nu '}$; note however, that its presence is not very
essential for the discuss of T-duality symmetry since the spacetime tensors
and coordinates are assumed be to inert under the T-duality transformations, 
as a consequence this term will be duality invariant on its own right.
Therefore, we shall deal with a single vertex function to discuss T-duality
symmetry as a prelude 
\bea
\label{vertexy}
V^{(1)}_1= A^{(1)}_{\alpha\beta ,\alpha '\beta '}(X)\partial Y^{\alpha}
\partial Y^{\beta}{\bar\partial}Y^{\alpha '}{\bar\partial}Y^{\beta '}
\eea
As argued earlier, if we expand out the expression for 
$V^{(1)}_1$, eq.(\ref{vertexy}),
 in terms of $P_{\alpha}$ and $Y'^{\alpha}$ we get terms of the following
type contacted with the tensor $ A^{(1)}_{\alpha\beta ,\alpha '\beta '}(X)$;
note that we do not use any symmetry(antisymmetry) properties of this tensor
under $\alpha\leftrightarrow\beta$ and $\alpha '\leftrightarrow\beta'$. 
Moreover, although we express the vertex in terms of $Y'$ and $P$, we still
like to retain the memory whether these terms came from left movers or right
movers. 
The full expression for the vertex function is  classified into five  
types. These are listed below:

\noindent (I) All are $P^{\alpha}$'s (index raised by $\delta ^{\alpha\beta}$):
$A^{(1)}_{\alpha\beta ,\alpha '\beta '}(X)P^{\alpha}P^{\beta}P^{\alpha '}
P^{\beta '}$.\\ (II) All are $Y'^{\alpha}$'s:
$A^{(1)}_{\alpha\beta ,\alpha '\beta '}(X)Y'^{\alpha}Y'^{\beta}
Y'^{\alpha '}Y'^{\beta '}$.\\
(III) The four terms with three $P^{\alpha}$'s;e.g
- $A^{(1)}_{\alpha\beta ,\alpha '\beta '}(X)P^{\alpha}P^{\beta}P^{\alpha '}
Y'^{\beta '}$.\\
(IV) There are also four terms with three $Y'^{\alpha}$'s which 
combine with the terms in (III) to give a T-duality invariant term; one such 
term is: 
-$A^{(1)}_{\alpha\beta ,\alpha '\beta '}(X)Y'^{\alpha}Y'^{\beta}P^{\alpha '}
Y'^{\beta '}$.\\
(V) There are six terms, each of  is a product of a pair of
momenta ($P^{\alpha}$ and a pair $Y'^{\alpha}$. A generic term is:
$A^{(1)}_{\alpha\beta ,\alpha \beta '}P^{\alpha}P^{\beta}Y'^{\alpha '}
Y'^{\beta '}$. 

\noindent We conclude from careful inspections of altogether 16 terms that, we
can combine terms in (I) and (II), those in (III) and (IV) and
those in (V) to compose $O(d,d) $ invariant functions. However,
 this is not an efficient method.\\
Let us consider the following three vertex functions in the present
compactification scheme
\bea
\label{v3}
 V^{(2)}_1= A^{(2)}_{\alpha\beta ,\alpha '}(X)
\partial Y^{\alpha}\partial Y^{\beta}{\bar\partial}^2Y^{\alpha '},~~~
 V^{(3)}_1= A^{(3)}_{\alpha , \alpha '\beta'}(X)
\partial ^2 Y^{\alpha}\partial Y^{\alpha '}\partial Y^{\beta '}
\eea
and
\bea
\label{v4}
V^{(4)}_1=A^{(4)}_{\alpha ,\alpha '}(X)\partial ^2Y^{\alpha}{\bar\partial}^2
Y^{\alpha '}
\eea
It follows from (\ref{relation}) that these vertex functions are related to
derivatives of  $A^{(1)}$ as a consequence of the constraints that they
be $(1,1)$ primaries. For the case in hand, when these carry all internal
indices and our focus is on (\ref{vertexy}), they will vanish.
Moreover, in order to study T-duality properties, of (\ref{v3}) and (\ref{v4})
we encounter another problem since higher order derivatives, 
$\partial, {\bar\partial}$ act on $Y$; consequently, the classification scheme
adopted to group terms as in (I)-(V), discussed above are not quite suitable. \\
Furthermore, we must recognize that, verifying T-duality symmetry for 
 higher excited states will provide obstacles since we have to deal with
product of a string of $\partial Y, {\bar\partial Y}$ and more and more
worldsheet derivatives acting on $Y^{\alpha}$ when we consider higher mass levels. 
One of the vertex functions for second massive level is
\bea
\label{secondmass}
V^{(1)}_2=C^{(1)}_{\alpha\beta\gamma ,\alpha '\beta '\gamma '}
\partial Y^{\alpha}\partial Y^{\beta} \partial Y^{\gamma}
{\bar\partial}Y^{\alpha '}{\bar\partial}Y^{\beta '}{\bar\partial}Y^{\gamma '}
\eea
expressed as products of $P^{\alpha}$, $Y'^{\alpha}$, $P^{\beta'}$ and 
$Y^{\beta'}$, all the terms can be reorganized and reexpressed in such a way 
that $V^{(1)}_2 $ is T-duality invariant. However, second level has  lot more 
terms and it is not easy to verify whether these vertex functions are
 T-duality invariant following the method alluded to above.\\
If one considers vertex operators for higher massive levels, the vertex operator
for each level is composed of large number of vertex functions. We propose the 
following  procedure to systematically organize various 
vertex functions at a given level.
(i) The first observation is that the basic
building blocks of vertex functions are 
$\partial Y^{\alpha}=P^{\alpha}+Y'^{\alpha}$ and 
${\bar\partial} Y^{\alpha '}=P^{\alpha '}-Y'^{\alpha '}$. 
(ii) Each vertex function 
at a given level is either string of products of these basic blocks or
these blocks are operated by $\partial$ and ${\bar\partial}$ respectively
so that each vertex operator at a given mass level has the desired dimensions.
Thus  it is not convenient to deal with $P^{\alpha}$ and $Y'^{\alpha}$
separately in order to study
the T-duality properties and the same is true for the
combinations $P\pm Y'$. However, $P^{\alpha}$ and $Y'^{\alpha}$ can be projected
out from the $O(d,d)$ vector, $\cal W$. \\
Let us first introduce following projection operators for later conveniences
\bea
\label{project1}
{\bf P}_{\pm}={1\over 2}({\bf 1}\pm {\tilde {\bf {\sigma _3}}}),~~~
{\tilde{\bf{\sigma _3}}}=\pmatrix{  1 & 0\cr 0 & - 1 \cr}
\eea
where $\bf 1$ is $2d\times 2d$ unit matrix and 
and the diagonal entries $( 1, -1)$ stand 
for $d\times d$ unit matrices. It is
easy to check that the projection operators are $O(d,d)$ matrices since each
one of them is. We project out two $O(d,d)$ vectors as follows
\bea
\label{vectorproject}
P={\bf P}_+{\cal W}, ~~~Y'={\bf P}_-{\cal W}
\eea
Therefore,
\bea
\label{py}
P+Y'={1\over 2}\bigg({\bf P}_+{\cal W}+ {\bf\eta}{\bf P}_-{\cal W}\bigg),~~
P-Y'={1\over 2}\bigg({\bf P}_+{\cal W}- {\bf\eta}{\bf P}_-{\cal W}\bigg)
\eea
notice that $\bf\eta$ flips lower component $Y'$ vector to an upper component 
one. Thus when we have only products  of $P+Y'$ and $P-Y'$, we can express them
first as products of $O(d,d)$ vector and subsequently contract their 
indices with appropriate tensors endowed with $O(d,d)$ indices.
Next we deal with worldsheet partial derivatives $\partial$ and $\bar\partial$
operating on basic building blocks. Let us define
\bea
\label{der}
\Delta _{\tau}={\bf P}_+\partial _{\tau}, ~~~
\Delta _{\sigma}={\bf P}_+{\partial}_{\sigma}~~~{\rm and}~~
{\Delta}_{\pm}(\tau,\sigma)={1\over 2}(\Delta _{\tau}\pm\Delta_{\sigma})
\eea
Therefore,
\bea
\label{pplusy}
\partial (P+Y')=\Delta_+({\tau,\sigma})\bigg({\bf P}_+
{\cal W}+ {\bf\eta}{\bf P}_-{\cal W}\bigg)
\eea
Thus the above expression is an $O(d,d)$ vector. Similarly, when $\bar\partial$
operates on $P-Y'$, we can express it as
\bea
\label{pminusy}
{\bar\partial}(P-Y')=\Delta _-({\tau,\sigma})
\bigg({\bf P}_+{\cal W}- {\bf\eta}{\bf P}_-{\cal W}\bigg)
\eea
The vertex operators we have considered in eqs.(\ref{vertexy}) and
(\ref{secondmass})
which are expressed as only string of products of $\partial Y^{\alpha}$
and ${\bar{\partial}}Y^{\alpha '}$ can be rewritten in terms of the $O(d,d)$ 
vectors $\cal W$ and subsequently contracted with suitable $O(d,d)$ tensors.
We remind the reader that, now familiar, $M$-matrix which expresses the
Hamiltonian in $O(d,d)$ invariant form  is also parametrized in
terms of backgrounds $G_{\alpha\beta}$ and $B_{\alpha\beta}$. Let us turn our
attentions to the other three vertex operators appearing in (\ref{v3}) and
(\ref{v4}). The procedure outlined above can be adopted to cast $V^{(2)}_1,
V^{(3)}_1$ and $V^{(4)}_1$ in a straight forward manner using the relations
(\ref{pplusy}) and ({\ref{pminusy}). \\
In order to illustrate the variety of vertex functions that can arise as we
go to higher levels; let us consider 
the second massive state as an example. We list below 
the vetex function which is  assumed to be sum of all vertex functions with 
{\it only} internal indices, momentarily assume all other vertex functions are
set to zero and this is the vertex operator. Thus vertex operator
 assumes the form
\cite{ov2}
\bea
\label{phitwo}
{{\bf{{\Phi}_2}}}= V^{(1)}_2+V^{(2)}_2+V^{(3)}_2+V^{(4)}_2+V^{(5)}_2+V^{(6)}_2
\eea
The vertex functions $V^{(i)}_2,i =1,6$ are given below 
(expression for $ V^{(1)}_2$ is given by (\ref{secondmass}).
\bea
\label{vtwo}
V^{(2)}_2=C^{(2)}_{\alpha\beta,\alpha '\beta '\gamma '}
{\partial}^2Y^{\alpha}\partial Y^{\beta}{\bar\partial}Y^{\alpha'}
{\bar\partial}Y^{\beta'}{\bar\partial}Y^{\gamma ' } +
C^{(3)}_{\alpha\beta\gamma,\alpha '\beta '}\partial Y^{\alpha}\partial Y^{\beta}
\partial Y^{\gamma}{{\bar\partial}^2}Y^{\alpha'}{\bar\partial}Y^{\beta'}
\eea
These vertex functions have a term of the form ${\partial}^2Y^{\alpha}$ or
${{\bar\partial}^2}Y^{\alpha'}$ and rest of the structure is decided by
dimensional considerations. Some of the other vertex functions are
\bea
\label{vthree}
&&V^{(3)}_2=C^{(4)}_{\alpha,\alpha'\beta'\gamma'}{\partial}^3Y^{\alpha}
Y^{\alpha'}
{\bar\partial}Y^{\beta'}{\bar\partial}Y^{\gamma ' } + 
C^{(5)}_{\alpha\beta\gamma,\alpha'}\partial Y^{\alpha}\partial Y^{\beta}
\partial Y^{\gamma}{{\bar\partial}^3}Y^{\alpha'},\nonumber\\
&&V^{(4)}_2=C^{(6)}_{\alpha\beta,\alpha'\beta'}{\partial}^2Y^{\alpha}
\partial Y^{\beta} {{\bar\partial}^2}Y^{\alpha'}{\bar\partial}Y^{\beta'}
\eea
and 
\bea
\label{vfive}
V^{(5)}_2=C^{(7)}_{\alpha\,\alpha'\beta'}{\partial}^3Y^{\alpha}
 {{\bar\partial}^2}Y^{\alpha'}{\bar\partial}Y^{\beta'}
+C^{(8)}_{\alpha\beta,\alpha'}{\partial}^2Y^{\alpha}\partial Y^{\beta} 
 {{\bar\partial}^3}Y^{\alpha'},~
V^{(6)}_2=C^{(9)}_{\alpha,\alpha'}{\partial}^3Y^{\alpha} 
{{\bar\partial}^3}Y^{\alpha'}
\eea
The tensors $C^{(2)}-C^{(9)}$ appearing in eqs.(\ref{vtwo}-\ref{vfive}) are 
all functions of $X^{\mu}$, independent of compact coordinates $Y^{\alpha}$,
and constrained by requirements of conformal invariance (not necessarily
nonvanishing in the compactification scheme we envisage). We observe from the 
structures of vertex functions $V^{(1)}_2 - V^{(6)}_2$ that, each one with
the combinations of the terms will be $O(d,d)$ invariant when we follow
the prescriptions of introducing projection operators, rewrite the
combinations  $P+Y'$ and $P-Y'$ as $O(d,d)$ vectors and 
convert $\partial$ and
$\bar\partial$ to ${\Delta_\pm(\tau,\sigma)}$ to operate on 
$P\pm Y'$ (reexpressed in terms of the projected ${\cal W}$'s)  
respectively. Let us consider $n^{th}$ excited massive level as an example.
The the dimension of all right movers obtained from products of 
$\partial Y$ higher powers of $\partial$ acting on $\partial Y$ should be
$(n+1)$ and same hold good for the left moving sector as well. Consider the
right moving sector of the type ${\bf \Pi}_1^{n+1}{\partial Y^{\alpha_i}}$ 
and the left moving sector ${\bf \Pi}_1^{n+1}{{\bar\partial} Y^{\alpha'_i}}$   
The vertex function is
\bea
\label{vertexn} 
V_{\alpha_1,\alpha_2...\alpha_{n+1},\alpha'_1\alpha'_2...\alpha'_{n+1}}(X)
{\Large \Pi}_1^{n+1}{\partial Y^{\alpha_i}}
{\Large \Pi}_1^{n+1}{{\bar\partial} Y^{\alpha'_i}}
\eea
and these products of ${\partial Y^{\alpha_i}}$ and 
${{\bar\partial} Y^{\alpha'_i}}$ can be converted to products of $(n+1)$ 
projected ${\cal W}$ for right movers and $(n+1)$ projected ${\cal W}$ from
left movers.
Let us consider  for a  vertex function for such a high level state.
A  generic vertex will have a structure
\bea
\label{generic}
{\partial}^pY^{\alpha_i}{\partial^q}Y^{\alpha_j}{\partial^r}
Y^{\alpha_k}...
{\bar\partial}^{p'}Y^{\alpha'_i}{\bar\partial}^{q'}
Y^{\alpha'_j}{{\bar\partial}}^{r'}Y^{\alpha'_k}...,~~p+q+r=n+1,~p'+q'+r'=n+1
\eea
The product is an $O(d,d)$ tensor whose rank is decided by the constraints on
sum of $p,q ~{\rm and}~ r$ and $p',q'~{\rm and}~ r'$ since number of 
$Y^{\alpha_i}$'s and $Y^{\alpha'_i}$'s appearing in (\ref{generic}) is 
determined from those conditions. Thus this tensor will be contracted with an
appropriate  tensor 
$T_{\alpha_i\alpha_j\alpha_k ..,\alpha'_i\alpha'_j\alpha'_k ..}(X)$ which will 
give us to a vertex function. Let us discuss how to express 
eq.(\ref{generic}) as a product of $O(d,d)$ vectors using the projection
operators introduced earlier.\\
(i) The first step is to rewrite\\
$\partial^pY=\partial^{p-1}(P+Y'),~~~~~{\bar\partial}^{p'}(P-Y')=
{\bar{\partial}}^{p'-1}(P-Y')$\\
(ii) We arrive at\\
$\partial^{p-1}(P+Y')={\Delta_+}^{p-1}(P+Y'),~~
{\bar{\partial}}^{p'-1}(P-Y')={\Delta_-}^{p'-1}(P-Y')$ \\
from (\ref{pplusy}) and (\ref{pminusy})\\
(iii) Finally, using the projection operators (\ref{py}) we get
\\
${\Delta_+}^{p-1}(P+Y')={\Delta_+}^{p-1}\bigg({\bf P}_+{\cal W}+{\bf\eta}
{\bf P}_-{\cal W}\bigg)$,~~
${\Delta}^{p'-1}(P-Y')={\Delta_-}^{p'-1}\bigg(({\bf P}_+{\cal W}-
{\bf\eta}{\bf P}_-{\cal W}\bigg)$
\\
Thus the products in (\ref{generic}) can be expressed as products of
$O(d,d)$ vectors. We need to contract these indices with suitable $O(d,d)$
tensors which have the following form:
\bea
\label{nlevel}
V_{n+1}={\cal A}_{klm..,k'l'm'..}{\Delta_+}^{p-1}{\cal W}_+^k
{\Delta}^{q-1}{\cal W}_+^l
{\Delta_+}^{r-1}{\cal W}_+^m..{\Delta_-}^{p'-1}{\cal W}_-^{k'}
{\Delta_-}^{p'-1} {\cal W}_-^{l'}
{\Delta_-}^{p'-1} {\cal W}_-^{m'}
\eea
where ${\cal W}_{\pm}=({\bf P}_+{\cal W}\pm{\bf\eta}{\bf P}_-{\cal W})$ with
$p+q+r=n+1$ and $p'+q'+r'=n+1$. 
Note that superscripts $\{k,l,m;k',l',m'\}$ appearing on ${\cal W}_{\pm}$ 
in eq. (\ref{nlevel}) are
the indices of the components of the  projected $O(d,d)$ vectors. Moreover, 
${\cal A}_{klm..,k'l'm'..}$ is $X$-dependent $O(d,d)$ tensor. Note that
(\ref{nlevel}) will be $O(d,d)$ invariant if coefficients transform
as follows \\
\bea
\label{rules}
{\cal A}_{klm..,k'l'm'..}\rightarrow \Omega_k^p  \Omega_l^q  \Omega_m^r ...
\Omega_{k'}^{p'} \Omega_{l'}^{q'} \Omega_{m'}^{r'}{\cal A}_{pqr..,p'q'r'..}
\eea
since each term in the product 
$\Delta ^{p-1}{\cal W}^k_+....\Delta^{p'-1}_-{\cal W}^{k'}$, above, transforms like
an $O(d,d)$ vector. 
It is worth while to dwell on another aspect of $O(d,d)$ invariant form of
the vertex function. For sake of definiteness focus on the vertex function
associated with the first massive level:
$A^{(1)}_{\alpha\beta,\alpha'\beta'}\partial Y^{\alpha}\partial Y^{\beta}
{\bar\partial} Y^{\alpha'}{\bar\partial} Y^{\beta'}$. We can always rewrite it
as
\bea
\label{irr}
T^{(1)}_{kl,k'l'}(X){\cal W}_+^k{\cal W}_+^l{\cal W}_-^{k'}{\cal W}_-^{l'}
\eea
We have distinguished the appearance of the $O(d,d)$ vectors in the
expressions for the vertex functions whether they originate  
 from right moving or left moving sector through unprimed and primed
indices. The vertex function (\ref{irr}) is $O(d,d)$ invariant; however,
it could be decomposed as sum of contracted tensors belonging to
irreducible representations of $O(d,d)$. \\
We first illustrate the point
by a simple example from atomic/nuclear physics when one considers
familiar multipole operators which usually appear in computations of
radiative transitions. The operator ${\bf{x^ix^j}}$ is decomposed into
\bea
\label{quadro}
{\bf x^ix^j}=({\bf{x^ix^j}}-{1\over 3}{\bf \delta^{ij}}{\bf x}^2)+
{1\over 3}{\bf x}^2{\bf{\delta^{ij}}}
\eea
Note that the first term is the quadrupole operator (traceless). If we
construct  a scalar ${\bf T_{ij}x^ix^j}$; the product decomposes into
${\bf{ T_{ij}Q^{ij}}}+{\bf T_i^i}{\bf x}^2$; ${\bf Q^{ij}}$ being the
quadrupole operator.\\
Let us examine the tensor structures in (\ref{irr}). $T^{(1)}_{kl,k'l'}$
is contracted with product of ${\cal W}_+^k{\cal W}_+^l$ and
 ${\cal W}_-^{k'}{\cal W}_-^{l'}$. Each of these tensors can be
decomposed as follows
\bea
\label{irrdd1}
{\cal W}_+^k{\cal W}_+^l &=&\bigg({\cal W}_+^k{\cal W}_+^l-{1\over{2d}}
{\bf \eta}^{kl}{\cal W}_+^m{\bf\eta}_{mn}{\cal W}_+^l\bigg)+
{1\over 2d}{\cal W}_+^m{\bf\eta}_{mn}{\cal W}_+^l 
\eea
and
\bea
\label{irrdd2}
{\cal W}_-^{k'}{\cal W}_-^{l'}=\bigg({\cal W}_-^{k'}{\cal W}_-^{l'}
-{1\over{2d}}{\bf \eta}^{k'l'}{\cal W}_-^{m'}{\bf\eta}_{m'n'}{\cal W}_-^{l'}
\bigg)+{\cal W}_-^{m'}{\bf\eta}_{m'n'}{\cal W}_-^{l'}
\eea
It is obvious that (\ref{irr}) will be composed of sum of terms arising from
IRR of $O(d,d)$,
taking into account
the decompositions (\ref{irrdd1}) and (\ref{irrdd2}),  which are generalization
of (\ref{quadro}) for the case at hand.  Therefore, a generic vertex function, 
for the
$n^{th}$ massive level, which assumes the
form (\ref{generic}), can be converted to an expression of the type
 (\ref{vertexn}) using our prescription. Since they are eventually expressed
as products of $O(d,d)$ vectors and contracted with suitable tensors.
These product of the the $O(d,d)$ vectors will be decomposed into direct
sums of the IRR of $O(d,d)$ and thus will contract with the decomposed
$O(d,d)$ tensors written as direct sums of IRR tensors. 
We conclude that all the vertex operators of each massive
level will be expressed as sums of IRR's of the T-duality group. Thus as we
go to higher and higher levels, we have to deal with higher and higher
dimensional representations of this noncompact group.\\
Now we turn our attention in another direction. Note that ${\bf{\Phi}_2}$ was
expressed as sum of vertex functions where tensors with only internal indices
were contracted with various types of derivatives of $Y^{\alpha}$. All these
levels are scalars under $ SO(D-1)$. However, once
we allow tensors appearing in vertex functions to carry Lorentz indices, these
tensors will be contracted with derivatives of $X^{\mu}$ and internal indices
will contract with derivatives of compact coordinates.
We claim that those vertex functions which have expressions with contraction of 
Lorentz indices with $X^{\mu}$'s will 
be $O(d,d)$ invariant with respect to rest of the tensor indices contracted 
with indices of $Y^{\alpha}$'s.
 Let us consider the first excited massive level to illustrate
our strategy which can be generalized to any level. We claim that full
vertex operator, for this level, are $O(d,d)$ invariant. We recall $X^{\mu}$
and tensors with only spacetime  indices (i.e. $\mu,\nu,..$etc.) 
a tensor transform trivially under the T-duality for these set of indices.
 Thus
\bea
\label{muindex}
{\tilde V}^{(1)}_1={\tilde A}^{(1)}_{\mu\nu,\mu'\nu'}\partial X^{\mu}
\partial X^{\nu}{\bar\partial} X^{\mu'}{\bar\partial} X^{\nu'}
\eea
is $O(d,d)$ invariant as per above prescription. Similarly, vertex function 
constructed out of :\\
 ${\tilde A}^{(2)}_{\mu,\mu'\nu'}\partial^2 X^{\mu}{\bar\partial} X^{\mu'}
{\bar\partial} X^{\nu'}$, ${\tilde A}^{(3)}_{\mu\nu,\mu'}\partial X^{\mu}
\partial X^{\nu}{\bar\partial}^2 X^{\mu'}$ and
${\tilde A}^{(3)}_{\mu,\mu'}\partial^2 X^{\mu}{\bar\partial}^2 X^{\mu'}$
are also $O(d,d)$ invariant. Let us classify the vertex functions according
to the spacetime and 'internal' indices they carry (with appropriate 
contractions of course).\\
(A) Vertex functions which have one Lorentz index and three internal indices:
\\
${\tilde B}^{(1)}_{\mu\alpha,\alpha'\beta'}\partial X^{\mu}\partial Y^{\alpha}
{\bar\partial}Y^{\alpha'}{\bar\partial}Y^{\beta'}+$  other terms by permuting
the indices.\\
(B) Vertex functions which have two Lorentz indices and two internal indices:\\
${\tilde B}^{(2)}_{\mu\beta,\mu'\beta'}\partial X^{\mu}\partial Y^{\beta}
{\bar\partial}X^{\mu'}{\bar\partial}Y^{\beta'}+$ other similar terms.\\
(C) Vertex functions with three Lorentz indices and one internal index:\\
${\tilde B}^{(3)}_{\mu\nu,\mu'\beta'}\partial X^{\mu}\partial X^{\nu}
{\bar\partial}X^{\mu'}{\bar\partial}^2Y^{\beta'}+$ other similar terms.\\
(D) Vertex functions of the type:\\
(i)$ {\tilde B}^{(4)}_{\mu\nu,\alpha'}\partial X^{\mu}\partial X^{\nu}
{\bar\partial}Y^{\beta'}+$  other similar terms.\\
(ii) ${\tilde B}^{(5)}_{\mu\beta,\alpha'}\partial X^{\mu}\partial Y^{\alpha}
{\bar\partial}^2Y^{\beta'}+$ other similar terms.\\
(iii) Vertex functions with second derivatives:\\
${\tilde B}^{(6)}_{\alpha,\mu'}\partial^2 Y^{\alpha}{\bar\partial}^2X^{\mu'}$
and ${\tilde B}^{(7)}_{\mu,\alpha'}\partial^2X^{\mu}{\bar\partial}^2Y^{\alpha'}
$\\
The vertex functions whose Lorentz index/indices are contracted with
$\partial X^{\mu},\partial^2 X^{\mu}$, 
${\bar\partial}X^{\mu'},\partial^2X^{\mu}$
will be inert under $O(d,d)$ rotations; however, rest of the indices
correspond to internal indices and those are contracted with 
$\partial Y^{\alpha}, {\bar\partial}Y^{\alpha'}, \partial^2 Y^{\alpha},
{\bar\partial}^2Y^{\alpha'}$ and so on. 
Moreover, the vertex functions considered above, (A)-(D), do not necessarily
vanish unlike the cases when some vertex function, carrying only internal
indices ($V^{(2)}_1$ - $V^{(4)}_1$), vanished as the consequences of conformal
invariance i.e. that these are $(1,1)$ primaries. This conclusion can be
easily verified from relations eqs. (\ref{relation}) and (\ref{gauge}).
We conclude that  only the worldsheet variables with internal indices,
such as $P\pm Y'$  are relevant
to construct $O(d,d)$ vectors which contract with corresponding indices of the
relevant tensors. 
We have laid down a procedures to construct $O(d,d)$
vectors from $\partial Y^{\alpha}, {\bar\partial}Y^{\alpha'}, 
\partial^2 Y^{\alpha},{\bar\partial}^2Y^{\alpha'}$ and other higher derivative
objects. For example,  
${\tilde B}^{(1)}_{\mu\alpha,\alpha'\beta'}\partial X^{\mu}\partial Y^{\alpha}
{\bar\partial}Y^{\alpha'}{\bar\partial}Y^{\beta'}$ has three 'internal'
indices of ${\tilde B}^{(1)}$ contracted with $\partial Y^{\alpha}
{\bar\partial}Y^{\alpha'}{\bar\partial}Y^{\beta'}$ and therefore, this
vertex function will be converted to an $O(d,d)$ invariant vertex function
which has a generic form  
\bea
\label{convert}
{\tilde T}^{(1)}_{k,k'l'}{\cal W}^k{\cal W}^{k'}{\cal W}^{l'}
\eea
This argument can be carried forward for all vertex 
operators at any massive level of the closed bosonic string. Moreover,
the type of vertex functions discussed in (A)-(D) correspond to massive
particles of various spins which fall into the representations of $SO(D-1)$.
Therefore, we are  able to conclude that vertex functions for massive
levels of a closed bosonic string can be cast in an $O(d,d)$ invariant form for
every level following the procedure presented here.\\
We have proposed a systematic procedure to obtain T-duality invariant
vertex operators for massive levels of a closed bosonic string when it is
compactified on $T^d$, the $d$-dimensional torus. It is assumed that the
tensor fields associated with these vertex operators depend only on the
spacetime coordinates, $X^{\mu}(\sigma,\tau)$ and are independent of the compact
coordinates, $Y^{\alpha}(\sigma,\tau)$. The duality invariance is manifest for
vertex operators of each level once one uses the projection technique to
convert $\{P,Y'\}$ to $O(d,d)$ vectors and/or their $\Delta_{\pm}$ derivatives.
These vertex operators can be expressed as sum of IRR of $O(d,d)$.\\
The T-duality symmetry plays an important role in string theory. We expect that
these symmetry properties will have important applications. Recall that
the T-duality symmetry has been widely applied to obtain new solutions to the
background configurations through judicious implementations of the solution
generating techniques. Thus given a configuration of  massive level 
background field it will be possible, in principle,  to generate another
background within the same massive level. Furthermore, there are evidences
that massive excited states are endowed with local symmetries. It is worth 
while to examine the implications of T-duality for those local symmetries.
\\
Another  point which deserves attention is to study the zero-norm 
states in this formulation.
It is well known that the existence of zero-norm states is quite essential 
in order that the bosonic string respects Lorentz invariance in critical
dimensions i.e. ${\hat D}=26$. This issue has been carefully analyzed in
\cite{ov2,lee}. We expect that these results will continue to hold good
when we are dealing with toroidally compactified closed bosonic string.\\
It is well known that very massive stringy states  possess exponential
degeneracy which has played crucial in deriving Bekenstein-Hawking entropy
relation for stringy back holes from the counting of microscopic states.
This high degree of degeneracy is also instrumental in deducing the 
thermal nature of emission spectrum of a stringy black hole. We expect that
some of supermassive states which also belong to the spectrum of the
compactfied string might exhibit symmetry properties which are yet to be
discovered.\\
In summary, we have investigated T-duality properties of the vertex
operators of excited massive closed string. We have proposed a prescription
to show that the vertex operators at every level can be expressed in 
manifestly duality invariant form according to the IRR of $O(d,d)$
group. These results might have important consequences to discover new
stringy symmetries.\\
It will be very interesting to
 implement the toroidal 
compactification procedure adopted in \cite{ms}   for the
vertex operators of excited states extending the method presented here. 
There are need  to improve this prescription considerably in order to overcome
some  difficulties. However, the toroidal compactification in its totality applied
to vertex operators through dimensional reduction is expected to unravel
more interesting features of T-duality in string theory.

\bigskip

\noindent {{\bf Acknowledgments}:} I am grateful to members of 
the String Theory Groups
at Institute of Physics and National Institute of Science Education and 
Research (NISER) for fruitful discussions. I am thankful to Anirban Basu for 
carefully
and critically reading the manuscript. This work was primarily supported by
the People of the Republic of India and partly by Indo-French Center for the 
Promotion of Advanced Research:
IFCPAR Project No.IFC/4104-2/2010/201.

\end{document}